# Multi-3D-Models Registration-Based Augmented Reality (AR) Instructions for Assembly

Seda Tuzun Canadinc*   Wei Yan†

Texas A&M University

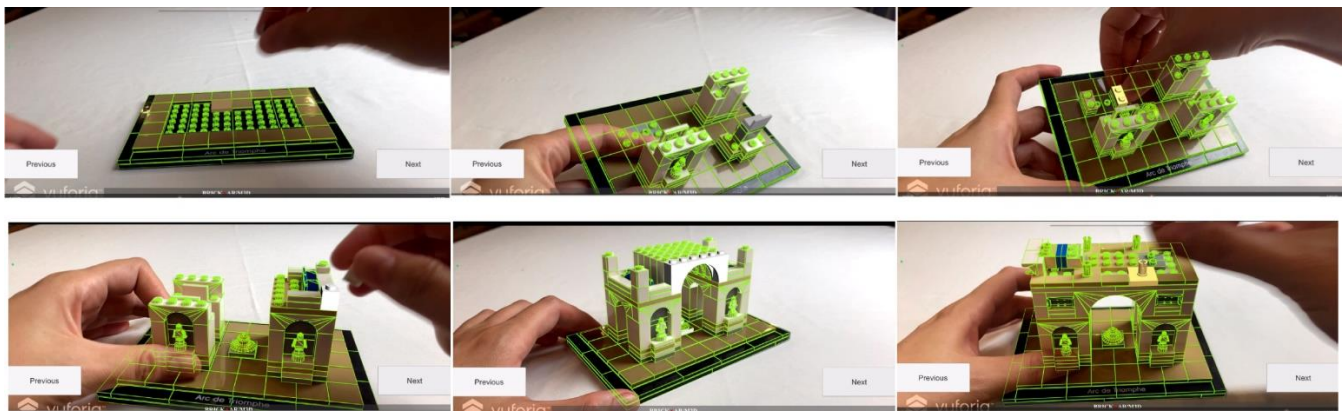

Figure 1: Scenes from BRICKxAR (M3D) at various steps with step numbers increasing from left to right.

**ABSTRACT**

This paper introduces a novel, markerless, step-by-step, in-situ 3D Augmented Reality (AR) instruction method and its application - BRICKxAR (Multi 3D Models/M3D) - for small parts assembly. BRICKxAR (M3D) realistically visualizes rendered 3D assembly parts at the assembly location of the physical assembly model (Figure 1). The user controls the assembly process through a user interface. BRICKxAR (M3D) utilizes deep learning-trained 3D model-based registration. Object recognition and tracking become challenging as the assembly model updates at each step. Additionally, not every part in a 3D assembly may be visible to the camera during the assembly. BRICKxAR (M3D) combines multiple assembly phases with a step count to address these challenges. Thus, using fewer phases simplifies the complex assembly process while step count facilitates accurate object recognition and precise visualization of each step. A testing and heuristic evaluation of the BRICKxAR (M3D) prototype and qualitative analysis were conducted with users and experts in visualization and human-computer interaction. Providing robust 3D AR instructions and allowing the handling of the assembly model, BRICKxAR (M3D) has the potential to be used at different scales ranging from manufacturing assembly to construction.

* e-mail address: sedatuzun@tamu.edu
† e-mail address: wyan@tamu.edu

**Index Terms**: Augmented Reality, AR, assembly guidance, in-situ AR, Human-Computer Interaction (HCI), AR-assisted assembly.

## 1 INTRODUCTION

There is ongoing research on Augmented Reality (AR) instructions for the parts assembly process in manufacturing, building construction, repair and maintenance, and Do-It-Yourself (DIY) furniture. While accuracy and precision are highly significant, the next most important feature is that the instructions are easy to understand [1]. However, achieving high precision and accuracy usually depends on additional external resources such as markers [2]. The use of markers may not be practical in some use cases and could spoil the immersive AR experience. The utilization of markers in assembly processes imposes limitations on the interaction with a physical model. These limitations arise from scenarios where the model obstructs the markers, where the model must remain on a surface, or when the markers are integrated into the model or move synchronously with the model. When the markers move with the model, at some point when the user rotates the model, the markers may become obstructed by the model, resulting in lost AR registration. If the user chooses to view the markers the whole time, they limit their ability to move the model and also miss out on the back views. Furthermore, integrating markers requires additional effort and can compromise the authentic appearance of the model; it may not be compatible with all shapes. Moreover, if the user cannot handle the model, then the instructions cannot be presented in AR at full potential, such as viewing the back of each item even though they are presented in 3D and juxtaposed in the assembly location. On the other hand, markerless AR options have not been able to accommodate the assembly processes or high precision. For instance, GPS-based AR does not work well inside buildings and is unsuitable for small-scale precision [3]. Although 3D-model-based AR promises high precision and tangibility [4], the model would change throughout an assembly and it would lead to the loss of AR registration since a previous phase of the same model, with a different shape, was used for the initial 3D-model-based AR registration.

To bridge these research gaps, BRICKxAR (M3D), a low-cost, immersive, tangible, mobile AR instructions framework for small parts assembly was conceived and a working prototype was developed that assures accuracy employing Multi-3D-Models Registration (M3D) AR to deal with the aforementioned problems. Our experiments using a LEGO model assembly and video recording of the experiments show that the prototype is working with a precise alignment [5].

1) How can one enable 3D AR assembly instructions while the physical 3D model is updated step-by-step?
2) How can one improve the design of the 3D AR assembly instructions toward optimized AR instructions?
3) What important new features should be developed to improve 3D AR assembly instructions?

In summary, we contribute the following:

- We developed a new method (BRICKxAR (M3D)) for 3D AR assembly instructions based on multi-3D-models registration that allows handling of the model during assembly.
- The 3D AR assembly instructions are organized and optimized considering AR model detection and tracking, structural stability, usability, and user experience.
- We performed testing of the working prototype and heuristic evaluations with users and experts to verify the proposed method and its usability.

## 2 RELATED WORK

Although the use of robots for the assembly of products is well-accepted in industry, manual assembly is still widely employed in many fields of manufacture such as aerospace, automotive, home appliance, and everyday objects industry, or other areas including building construction, repair, maintenance, inspection, DIY furniture, training, and education. For education and training, especially in architectural and engineering fields, an assembly can be at full scale through design-build projects [6]. Assembly can also be a reduced scale model yet significantly contribute to the understanding of structures or details. With the introduction of the Model-Based Definition (MBD) approach, 3D digital model data replaces traditional 2D drawings in assembly instruction booklets [7]. Studies show that along with planning and presentation [8], ease of assembly significantly contributes to consumers' overall satisfaction with ready-to-assemble furniture [9] and users' performance, experience, and learning in work or education-related assembly tasks.

### 2.1 Traditional Guidance in Assembly

Usually, in traditional assembly instructions on paper, 3D model data are provided in an exploded view, enriched with text and action diagrams, following a hierarchy of subassemblies and step-by-step instructions [8]. For instance, to make it simple and universal, furniture assembly booklets provide generalized templates with pictorial instructions rather than text in any specific language to demonstrate the steps involved in assembly [9]. Users are required to synthesize the 2D isometric drawings in the traditional instructions into 3D mental representations [10]. Since cognitive resources are limited even for trained users, the cognitive load required for each instruction and task performance may cause mental fatigue [11]. Mental fatigue may lead to errors, especially in complex assembly processes [12].

### 2.2 AR-Assisted Assembly

On the other hand, an AR-assisted assembly is an approach that combines AR technology with the process of assembling objects. AR can not only present virtual objects in physical environments but also align physical objects and virtual ones in real time [13]. Moreover, AR allows immersive experience and interaction [14]. Also, it bridges the gap between the virtual world and the physical world, both spatially and cognitively [15]. As a solution to the increasing complexity of tasks in manufacturing and construction, knowledge-based AR is developed. Knowledge-based AR assists users so that errors are reduced and long training is eliminated. Besides, AR adoption enables human-robot collaboration in manufacturing [16]. A study shows that assembly tasks constitute 18% of AR applications in the construction phase of projects [12]. 60% of AR use depends on marker-based AR, while the rest of AR use is markerless, employing a projection-based approach, the Global Positioning System (GPS), or Radio Frequency Identification (RFID smart labels) for localization [12]. However, during marker or RFID tag usage, the location of the marker/tag should be stable, and also, the marker/tag placement in relation to the model should remain the same during assembly; otherwise; misplacement would cause errors. Moreover, GPS does not work accurately indoors and requires adjustments to precisely align content inside the buildings [3].

### 2.3 Benefits and Impacts of AR-Assisted Assembly Instructions

AR assembly instructions present 3D spatiality which the traditional 2D instructions lack, and can employ motion of artifacts such as rotating each piece to properly display them in all views and make every detail visible to the user and this facilitates part identification. Besides, AR instructions can display additional visual data animation such as arrows pointing to the assembly location. In addition, audio information about each piece could help part identification, especially for special needs people, this may provide many advantages. With markerless AR instructions, the order of instructions gets free from the restrictions of booklet instructions that limit the views to present 3D objects in 2D and the need to stick to the same view angle most time. Due to the separation of instructions and the assembly in two different locations, paper-based instructions require users to switch their focus between them. However, in-situ AR instructions have the potential to alleviate this cognitive load. There are studies employing markerless object tracking to provide both tangibility and in-situ AR content [17], in which a user can handle the model, and see and understand 3D spatial relationships between pieces. These advancements are expected to reduce errors, cognitive load, and task completion time. Besides, part identification in AR can be automated using Machine Learning (ML) (computer vision) for object recognition. For instance, Brickit is a free mobile application, and with its ML integration, it scans a pile of LEGO bricks and identifies every brick in the pile [18]. The app matches your pile of bricks with its assembly instructions inventory and shares with you the possible assembly instructions digitally as 2D that you can build with what you have in your pile [18]. Funk et al. present an AR instructions method including an augmented workplace for part identification; the assembly is conducted with LEGO Duplo bricks as a representation of an industrial assembly task and bricks are collected in separate bins according to their types and colors and the bins' locations are preorganized [19], [20].

### 2.4 AR-Assisted Assembly in Manufacturing and Architecture, Engineering, and Construction (AEC)

Products have become more varied, complex, and complicated because of the customization of products to meet the needs of each customer [21]. Assembly plays a vital role in fast response to customization with personalized modules sub-systems [22]. The latest trends in globalized manufacturing demand real-time information exchanges between several product development stages including assembly [22]. Complexity refers to being made up of many interconnecting parts [23] which we see ubiquitously in manufacturing and construction assemblies. Dynamic environments and demanding time limits in industry and construction make them risky and complex businesses with a high number of uncertainties [23]. According to one study, project complexity may contain one or more of 5 main aspects: (1) Organizational (people involved/relationships), (2) Operational and technological, (3) Planning and management, (4) Environmental, and (5) Uncertainty [23]. AR has become a human-computer interaction (HCI) technique that has the potential to address these problems by overlaying digital objects and content on top of the physical environments [24].

The implementation of AR-Assisted assembly is limited in the architectural field at the moment [25], AR-related research conducted in the last ten years focuses on three major categories: 3D AR assembly instructions, AR data sharing, and AR for HCI [26]. Introduction of Computer-Aided-Design (CAD) software tools especially paired with visual programming (Rhino-Grasshopper, or Revit-Dynamo), enables architects to create complex architectural forms. Besides, AR tools enhance students' understanding of spatial transformations and their mathematical representations that are essential in CAD [27]. However, physically building these complex forms is not always straightforward and requires skilled workmanship. AR-Assistance could help build those complex forms with the guidance of 2D or 3D instructions displayed on AR devices even without professional workmanship knowledge or experience. For instance, Jahn et al. used AR-Assisted assembly to demonstrate the approximate simulation of steam-bent timber's material behavior in the design and construction of a pavilion [28].

With the advancement in AR, robotics, and 3D printing technologies both in manufacturing and construction, there is a need to redesign assembly processes together with operation facilitation [24]. Furthermore, it's important to consider human-robot collaboration as a significant component of this process [29] [30]. Another aspect is the occupational health and safety issues in assemblies in which operators are involved. To minimize risks and meet deadlines, AR-Assisted assembly instructions can help simplify the complexity of assemblies, reduce errors, and prevent injuries [31].

## 3 METHODOLOGY

One of the main objectives of this research is to explore and develop methods that enable the use of markerless 3D augmented reality (AR) assembly instructions while the physical 3D model is updated step-by-step. After the completion of each assembly step, the modified model would no longer match the 3D object recognition model, eventually resulting in a loss of registration. On the other hand, using object recognition for each step would be challenging for a few reasons:

1) Object recognition is not as sensitive as detecting small part additions.

2) Some parts may be obstructed by other parts due to the geometry of the model, which could potentially be a hurdle for object recognition.

It may be challenging to provide an ideal viewing angle for each step to correctly register for the exact step. Otherwise, object recognition could get confused to register for the correct step. Providing an ideal viewing at each step would be burdensome for the user.

### 3.1 Assembly Phases

To overcome the challenges of the current 3D model-based object recognition problems, the use of assembly phases was hypothesized. Instead of object recognition for each step, the whole assembly was divided into phases, and deep learning training was conducted for object detection in each phase. Since object recognition has a tolerance, it would stay registered after its registration for several steps until the next phase. Also, if two deep-learning trained object recognition models are close to each other in geometry, that would lead to errors in recognition due to the tolerance. So, in the prototype, phase recognition was paired with a separate step count to control object recognition and registration at certain steps. This enabled both accurate model registration of geometric transformations (location and orientation) and part visualization of the correct step.

### 3.2 Selection of Development Tools

For the AR software and deep learning training for object recognition of the LEGO model phases, Vuforia which is an AR software development kit (SDK) [32] was used. Deep learning training was implemented in Vuforia's standalone tool, Model Target Generator (MTG). In MTG, deep learning training was conducted by providing a CAD model in obj format. MTG allows for automatic color assignment for each part to potentially enhance object recognition. Textures, patterns, and non-flexible, rigid geometry with sufficient visual features and details also facilitate object recognition. The deep learning training process is conducted using the resources in the Vuforia Cloud. In Vuforia Advanced Model Target, the CAD model is trained to be recognized by the AR app. The CAD model is run through a million random background images, different lighting conditions, poses, or angles with different colorizations. In the end, a neural network file or database is created. This database can be put in the app and allows that object to be recognized based on the object's shape or edges [33], [34].

### 3.3 Selection of Test Case

As a test case for the prototype, the LEGO® Arc de Triomphe set from the LEGO® Architecture theme [35] was chosen. It is a physical model with 386 parts. The CAD model of the LEGO Arc de Triomphe model was also procured in the formats of fbx and obj. However, The LEGO Arc de Triomphe model is challenging for AR testing due to its symmetry. 3D-model-based AR registration works well with asymmetrical models because they provide different edges on both sides which ensures robust registration, but symmetrical models may lead to confusion due to ambiguous orientations and brick locations. The choice of a symmetrical model was intentional to investigate challenges in the planning of the BRICKxAR (M3D) phases.

### 3.4 Assembly Sequence Planning and Phasing

Experimentation with object recognition aided the estimation of sensitivity and the factors affecting it. Since the parts are tiny, the registration is maintained even though several parts are added. The size of the bricks, their location, and their effect on edge detection contribute to object recognition. The order of the assembly, orientation, user experience, structural stability, and limiting the number of phases for reducing the required deep learning training models must all be considered. AR instructions do not necessarily need to follow the same order as paper booklet instructions and can be more flexible. Additional considerations are subsequent assembly steps, locating a sequence of operations in proximity, maintaining, structural stability, providing a strong base for subsequent steps, and assuring attachment under pushing, pulling, and holding forces. Because object recognition is not effective with a rectangular brick or a few bricks due to a lack of sufficient features and details to detect, subassemblies were eliminated to avoid the need to start a new sequence. In addition, providing correct orientation during assembly is another significant factor that affects the assembly order. To accurately register the virtual model on top of the physical model, there is a need to have a correct sense of orientation about which side is the front, back, right, and left. To provide an orientation to the model, the phase models were optimized with asymmetrical features and edges confirmed empirically by experiments.

Eventually, the assembly process was divided into five phases:

In Phase I, ground plane registration was implemented by using a flat surface to anchor the virtual assembly. Because in the beginning, there was no model to be detected, which was needed to start the 3D model-based registration. Phase I used Ground Plane Detection to ensure a correct first step. Phase I presented step-by-step 3D AR instructions registered on the anchored surface location up to step 104.

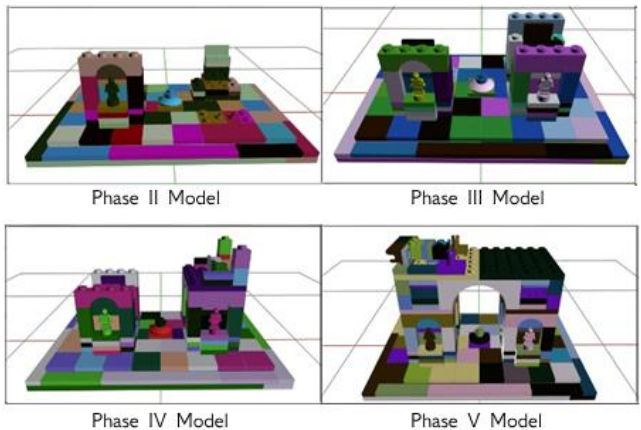

Figure 2: BRICKxAR (M3D) Assembly Phase Models

Each CAD model of the rest of the four phases was devised through experiments to optimize the assembly order to consider several different factors affecting object recognition (distinct edges and orientation), structural stability, and user experience. The screenshots of the models from the MTG for each phase (Phase II-III-IV-V) are seen in Figure 2. Also, coloring the models diagrammatically for each phase model was chosen over the realistic appearance because it enhances model tracking performance and robustness [36]. Colors were assigned to the models by the MTG automatically. An Advanced Model Target database was generated for each phase individually.

### 3.5 Refinement of the Prototype

Prototyping was conducted using the following software tools: Vuforia Model Target Generator 9.4.6 (MTG) was used for deep learning training for object recognition[37]. Vuforia Engine Package for Unity 9.8.8 was used for AR development in Unity [38]. Unity 2020.3.13f1 was used for combining datasets from MTG, rendered CAD models of different phases of the LEGO model, Vuforia AR camera, and model target C# scripts[39]. Visual Studio 8.10.7 (build 17) was used for C# scripting and transparent and wireframe shaders for object occlusion and helping visualize instructions [40].

For the presentation part, BRICKxAR patented technology [2] for object occlusion is employed which includes Phantom rendering [41] and wireframe visualization that provides visual guidance of assembly steps through part visibility and realistic visualization of object occlusion. For each phase, three models have been juxtaposed: the rendered model, the wireframe model, and the transparent model (Figure 3). For a realistic object occlusion, the transparent model represents the physical model (LEGO bricks) in the assembly, so that in the AR app physical bricks occlude the virtual ones based on their spatial relations.

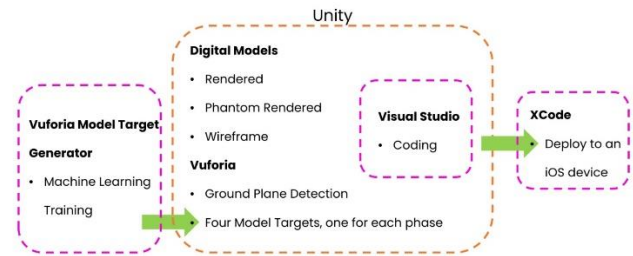

Figure 3: The structure of BRICKxAR (M3D) prototype.

The requirements for a working 3D model-based object recognition and AR registration for a 3D AR small-part assembly instruction are:

1) A physical model and a matching CAD model;
2) An asymmetrical assembly model;
3) An assembly model with distinct features to enable robust object recognition;
4) Tackling with the tolerance of 3D model-based object recognition;
5) Tackling with the start of the assembly process that does not have a 3D model with distinct edge features in the beginning.

One major challenge is starting the prototype because there is no physical model available at the beginning of the assembly process. Experiments have been conducted to use the first brick as the first model registration phase model and the first couple of bricks which form an asymmetrical edge. However, the geometry of a single rectangular brick and the geometry of the first few bricks do not allow a robust 3D-model-based AR registration. So, Ground Plane detection is implemented for starting the assembly process (Figure 4). The Ground Plane method starts with detecting a surface and displays the registered surface.

When the step count = 105 (for the LEGO Arc de Triomphe example), the ground plane detection is disabled and the ground plane detection models (the rendered, the occlusion, and the wireframe) become invisible. Simultaneously, the Phase II model target starts tracking, and the wireframe model of the parts from the previous steps become visible (but can be hidden if desired), while the part with index number 105 becomes rendered. The current rendered part and the previous parts are juxtaposed on top of the physical LEGO model. This allows the user to handle the model while the 3D AR instructions would move along with it. The similar AR instructions continue in the following steps. At step 129, there is a switch from the Phase II model to the Phase III model. To ensure a seamless transition, one step before the switching step (at step 128), the dataset belonging to the next phase (Phase III) is activated. At step 165, the Phase III model target switches to Phase IV, and at step 226, the Phase IV model target switches to Phase V (Figure 4).

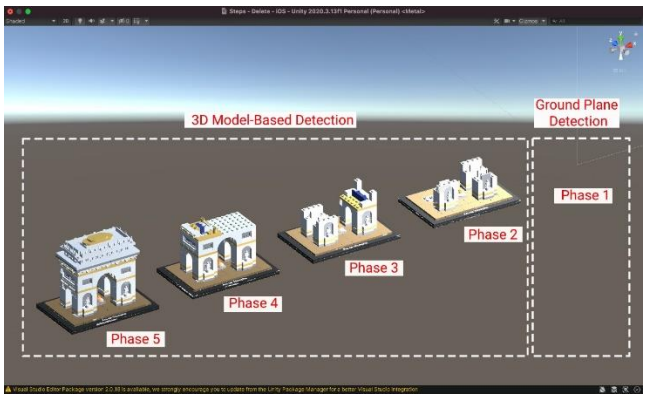

Figure 4: Different phase models in Unity scene.

The developed AR app is deployed to an iOS mobile device using Xcode. The versions of the software are as follows: Unity 2020.3.13.f1, Vuforia Engine Package for Unity 9.8.8, Vuforia Model Target Generator 9.4.6, Visual Studio Mac 8.10.7, Xcode 12.4; and iOS 14.6; however, Xcode 12.4 is modified to support iOS 14.6. The prototype is stored in GitHub.

## 3.6 Testing

Evaluation and validation of the research followed a mixed-mode method:

1) Experimental research through testing of the prototype conducted by the author by observing and making quantitative measurements of AR registration on top of the 3D model at various moments during sessions (Ground Plane registration and Model Target registration phases);

2) Heuristic evaluation and user testing of the proposed AR instruction and its design by UI experts and builders with an assembly task;

3) Qualitative analyses of discovery interviews.

### 3.6.1 Experimental Research

For the first part of the evaluation, the whole assembly process of 386 LEGO bricks was recorded using a screen recorder on a mobile device. An iPhone 11 with a screen resolution of 1792x828 pixels was used. The whole process was observed in terms of registration stability and accuracy, the transition quality between different 3D-model-based AR registration phases, and hand occlusion impact on the registration. Selected screenshots of different phases were analyzed in Adobe Photoshop. These screenshots were used to measure the errors between the digital parts added in AR and the actual physical model in Adobe Photoshop. The screenshots were evaluated by drawing perpendicular lines between the edges of the physical LEGO bricks and the AR wireframe lines. The gap between these two edges is measured by the distance of the in-between perpendicular line added to the images.

The findings show that Model Target registration in all phases is consistently accurate (Figure 5), but Ground Plane registration varies. Ground Plane relies on screen tapping for scale, making it hard to align digital and physical models. Users need trial and error to match the scale and position the physical model correctly.

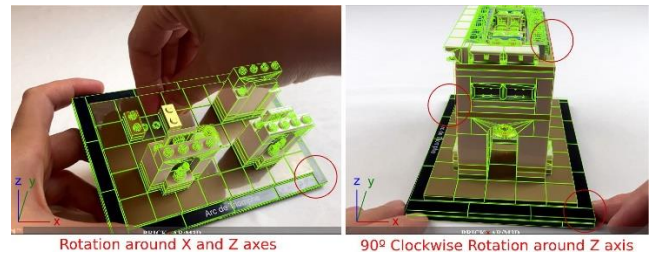

Figure 5: In BRICKxAR (M3D) when the model is rotated around both X and Z axes, the error (for alignment of physical and virtual brick edges) is less than 1 pixel shown in red circles. (XYZ axes and red circles were added later in Adobe Photoshop.)

Model Target registration in 360 degrees works best with the LEGO bricks model's front view, requiring slow movements for stable AR registration. Switching between Ground Plane registration and BRICKxAR (M3D) Phase I is seamless as well as switches between the other phases, also during disassembly. Even when both hands obscure over 66% of the model, the registration remains stable (Figure). Rotating the model around the X and Z axes minimizes errors, but exceeding a 90º rotation around the Z-axis may cause issues (Figure 5). Fast movements, occlusion, and camera focus may affect AR registration and can lead to flickering and degradation.

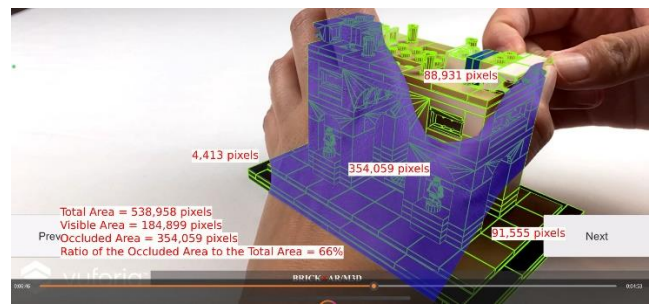

Figure 6: BRICKxAR (M3D) occlusion test. The image is a proportionally scaled screenshot of an original AR scene video image.

### 3.6.2 Heuristic Evaluation Study

To evaluate the usability and design of the BRICKxAR (M3D) prototype, 30 builders and 3 expert faculty members with expertise in Visualization and Human-Computer Interaction were interviewed. They were asked to perform the following:

- Answer questions about assembly, assembly instructions, and AR,
- Watch a video of a marker-based version (BRICKxAR [2]),
- Perform a simple LEGO assembly task of 10 parts while guided by the BRICKxAR (M3D) prototype (Figure 7),
- Conduct heuristic evaluations of the BRICKxAR (M3D) and its unique features,
- Compare the prototype with the marker-based version.

Heuristic evaluation is a method for detecting user interface usability problems conducted by a small number of evaluators validating user interface following the heuristics (usability principles) [42]. Heuristic means a rule of thumb and design heuristics are guidelines for the design and evaluation of a system's usability [43]. For this research, the heuristic evaluation followed the guidelines of AR design and design principles of assembly instructions recommended by the literature [2]. The heuristic evaluations were conducted following these principles of usability and design heuristics:

- current, previous, and future parts visibilities, aligning physical and virtual content and minimizing cognitive loads,
- unrestrictive and 360º viewing, rotation and zoom options,
- 1:1 scale and automatic player positioning,
- minimal looking at the instructions,
- comparison of consecutive structural diagrams,
- virtual cutaways,
- accessibility of offscreen objects,
- match with the hardware platform in terms of capabilities and limitations,
- eliminating repeated actions, and
- getting trained from the booklet drawings [2].

The planning of the design heuristics study and the interview questions are shown in Table 1.

Table 1. Design Heuristic Evaluation User Study Questions

| Design Heuristics Evaluation Interview Questions |
| --- |
| How often do you assemble a LEGO set or bricks? |
| How do you feel when assembling? What's your experience like? |
| What is the hardest part? Pains vs. Gains? |
| What do you use as assembly instructions? |
| How could instructions be made easier/better? |
| What do you like best/worst in the instructions? |
| How do you solve problems during assembly? Alternative ways? |
| How do the price of LEGO sets or bricks affect you? What is a complex build for you (how many pieces)? |
| Have you heard about AR instructions? Mobile AR? |
| What would be the impact of AR instructions? |
| Would AR instructions add value and why? |
| What do you think about AR instructions value about engagement and educational assistance? |
| What are your concerns about AR instructions? (Screen time vs. Play time) |
| Where do you learn about new technology? |
| Do you buy apps? Would you consider buying a LEGO set with AR instructions even if you need to pay a little more than a set itself? |
| AR instructions, nice to have or must have? |
| What do you think about AR helping children to be independent? |
| Are you willing to wear AR glasses throughout the assembly? |
| What do you think about/How do you compare the value of AR instructions on mobile versus AR glasses? |
| BRICKxAR Video Demonstration (Marker-based 3D AR instructions for assembly)[2] |
| Assembly task of 10 parts in the BRICKxAR (M3D) prototype |
| While using paper booklet or pdf, you switch between the assembly and the instructions. However, markerless AR instructions are juxtaposed. How do you compare the two? |
| Markerless AR eliminates subassemblies. What do you think about it? How do you compare assembling to the model to preparing subassembly? |
| Marker-based AR in the Video on the table. Markerless, you hold it. How do you compare them in terms of assembly & detecting errors? |
| Players can hold LEGO brick model in hand, rotate it in all directions, and assemble new pieces – What do you think about BRICKxAR (M3D) help you detect errors compared to using image markers as shown in the video? |
| Have you experienced any problems during transitions in the assembly? Is the multiple phase transition smooth enough so that the players have no problem at the transitions? |
| Is the asymmetric model assembly process acceptable? Explain this to the users. |
| How important does holding the LEGO brick model in hand compared to putting it on the table in terms of understanding spatial relations, visibility of the parts, and possibly learning spatial transformations? |
| How do you rate the efficiency of the app usability? 1-10 |

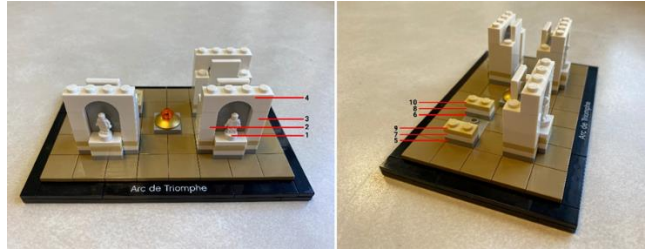

Figure 7: Assembly task: The location and order of the parts.

During the evaluation, users assembled 10 parts, with four in the front left tower and six hidden at the back, challenging visibility when viewed from the front (Figure ). Additionally, the model transitioned between phases during assembly, testing multi-model registration and phase transitions. Users were provided with a hands-free tripod and only necessary parts to simplify the task (Figure ). They were guided on using "Next" and "Previous" buttons and encouraged to handle the model. The study emphasized the importance of good lighting, with testing primarily conducted in well-lit, daylight environments to avoid AR registration failures in poorly lit spaces.

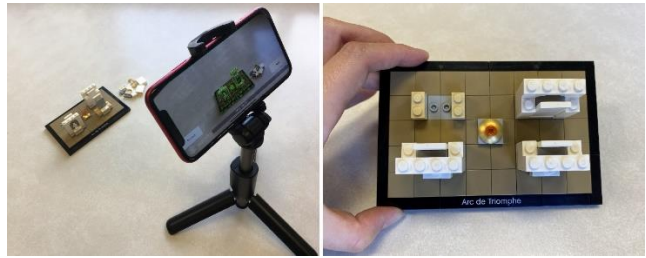

Figure 8: Use of a tripod in the builder heuristic evaluations and handling of the model.

### 3.6.3 Results

Among the 33 participants of heuristic evaluation, a demographic breakdown revealed that 46% were female and 55% were male, with actual numbers of 15 female and 18 male participants. During the assembly, most users view the mobile phone screen, then switch their gaze to the assembly to see the actual assembly with their eyes. Also, most of them are not willing to handle the model. Even if they need to handle the model to see the assembly locations at the back, they preferred rotating the model that sits on the table instead of handling and rotating the model in hand. One of the users reports that handling is good for visualization; however, the model on the table is good for assembling. Another user says it is better to have the model on the table because it frees one of your hands for holding the device, but AR goggles would be better to support a handling option. One of the users who is an advanced LEGO brick builder who assembles complex models, emphasizes holding the model would only work for small sets, and complex sets with more than a thousand parts are too large to handle them and rotate them in hand during the assembly.

One of the lessons from testing the prototype is the significance of the rendering scene in Unity where the lighting is placed to make the rendering more realistic and the virtual objects well-lit. Some users have difficulty identifying the color of the bricks at the back location which is colored gray, white, and yellow and that led them to make errors in the assembly order of the bricks.

Most users could not notice where the next part goes when the assembly part shifted to the back of the model in an unseen location. Most of them state that it would be helpful if there is an instruction or a visual clue about this to guide them to the exact location.

Some users think marker-based BRICKxAR and markerless BRICKxAR (M3D) are very similar to each other. Most users do not think changing the order of the assembly to optimize it and breaking the symmetry during the assembly makes a difference. Similarly, they think eliminating subassemblies do not make any difference.

One of the experts reports that using AR goggles during long hours of assembly may be tiresome because of the headset's weight and the position of the head putting all the weight of the headset on the neck while looking downwards at the assembly. Also, he notes that if the AR instructions are on the mobile there will be two different views: that the view of the user, and the view of the mobile, and he adds they should be aligned for AR to be more realistic.

## 4 CONCLUSION

The contributions of this research are as follows:

1) 3D model-based in-situ step-by-step 3D AR instructions;
2) The design of the assembly order;
3) Ways to improve 3D AR instructions.

The major contribution is presenting a method for 3D model-based in-situ AR instructions while the model is updated step-by-step. The developed method combines phases of assembly with a step count. Thus, using fewer phases simplifies the complex assembly process while step count facilitates accurate object recognition and accurate visualization of the correct step.

Another contribution is presenting ways to improve the design of the 3D AR assembly instructions toward optimized AR instructions. The assembly order is reorganized according to the requirements of AR registration. To ensure a robust AR registration, the physical model needs to have an asymmetrical form with distinct edges. Besides, there are other things to consider in designing the assembly order such as structural stability, usability, and user experience. The research provides an optimized assembly order that satisfies those needs.

The last contribution is providing what new features should be developed to improve 3D AR assembly instructions to better function and facilitate usability.

This research shows using multiple models for multiple phases of assembly for object tracking and registration paired with step count provides an effective way to guide users with 3D AR instructions and allows users to interact with models both physically and virtually while executing step-by-step assembly. With this methodology, these challenges are overcome:

1) Model-based registration's insufficient sensitivity to recognize each step and tolerate changes, especially with small parts;
2) Recognition of assembly steps in locations occluded by other parts of the model;
3) Recognition of assembly steps in locations that do not have any effect on the edges (because for accurate edge detection, edges should provide a distinct boundary);
4) Limitation of a model to stay on a surface during an assembly (BRICKxAR (M3D) allows for handling the model);
5) Limitation of a model's movements not to occlude a marker or markers (BRICKxAR (M3D) allows for moving and rotating the model).

Experiments with object recognition show that there is a tolerance with the detection and small changes such as additional small parts although changing the edges of a model may not prevent the registration. Based on experiments with certain assumptions (due to lack of documentation of Vuforia Model Target's detailed object tracking method) the factors affecting registration in the BRICKxAR (M3D) prototype lie in seven categories: (1) Part (Brick) Size, (2) Part Geometry, (3) Assembly Location, (3) Assembly Location, (5) Ratio of Part Size to Model Size, (6) The number of assembly parts, (7) Assembly Order. The parameters that affect the last category assembly order are: i) Object recognition, and edge detection, ii) Structural stability, iii) User experience, iv) Efficient planning and time considerations, v) The multi-objective optimization of these factors above to ensure the most efficient optimum assembly order.

Furthermore, user studies and heuristic evaluations show that model phase transitions mostly work well if there is good lighting. Text or audio guidance for identifying parts and visual guidance for identifying the assembly location would be helpful. Moreover, guidance needs to be improved especially when an assembly location is not visible to the user.

### 4.1 Limitations

The prototype requiring the mobile device to work becomes a barrier for some of the users, so most users use the screen for instructions and when assembling they turn their gaze to the assembly outside of the screen. This limitation could be overcome by using AR goggles/headsets; however, AR headsets are far from affordable and the technology is not ready for widespread commercialization yet.

Another limitation is determining the phase models by a trial-and-error method which is not efficient and takes time. Also, preparing instructions is made for each model individually. Besides, fast movements cause loss of registration during assembly and lack of sufficient lighting prevents registration from working accurately. Another limitation related to lighting is the scene lighting of the models in Unity which leads to confusion in color identification of parts.

A potential limitation that is not related to specific models' geometry is hand occlusion during the assembly. Furthermore, at the start of the application, we needed to implement a method other than object recognition since there is no model in the beginning.

Ground plane registration is used to provide the first phase model; however, in ground plane registration, the guidelines of the rectangular markers for tapping shown on the screen is not consistently displayed at the same size each time.

One of the challenges noted by participants in user study is error detection. The current BRICKxAR (M3D) prototype does not have automatic error detection; however, considering the tolerance of model-based registration and its functioning principles based on edges, color, pattern, and texture, the automatic error detection is not implemented for now.

## 4.2 Future Work

The findings from the evaluations, user tests, and heuristic evaluations give clues toward future work for the developed prototype. To improve the BRICKxAR (M3D) prototype to fit user expectations better, some features are needed to be added and new technology is needed to be developed and integrated:

1) **Brick Identifying and Error Finding** using Convolutional Neural Network (CNN) trained on available datasets or future developed datasets.

2) **Assembly Assistance** with the integration of audio and text notifications when the assembly location is hidden or partially blocked from the view of the user.

3) **Assembly Order Optimization** with the help of AI tools (large language models) to develop a multi-objective optimization algorithm to optimize phase models by taking into consideration various challenges, including but not limited to structural stability, user experience, and parts visibility.

4) **Assembly Instructions Automatization** with the integration of large language models for automating the planning of 3D AR instructions and auto-creating the instructions for a step-by-step assembly with notifications and audio helpers to guide users throughout an assembly.

5) **Subassemblies** could be defined as an object for object recognition and images of subassemblies can be used in deep learning training datasets.

6) **AR Goggles** will be used for future assembly instructions for hands free experiences.

7) **Collaboration** could be implemented to enable human-computer interaction and humans and robotic arms could collaborate through AR in the assembly.

## Acknowledgments

The research is funded by the National Science Foundation (NSF), I-Corps Grant No. 2242432, and partially funded by the NSF RETTL Grant No. 2119549, and Texas A&M University's Translational Investment Fund (TIF) from the Texas A&M Innovation, Mattia Flabiano III AIA/Page Southerland Design Professorship, Department of Architecture's Instructor of Record and Teaching Assistantship fund, Lechner Scholarship from Texas A&M University Department of Architecture, and Edward J. Romieniec Graduate Travel Fellowship from Texas A&M University School of Architecture.